# Speeding simulation analysis up with `yt` and Intel Distribution for Python

*Salvatore Cielo, PhD, scientific computing expert, Leibniz Supercomputing Centre*
*Luigi Iapichino, PhD, scientific computing expert, Leibniz Supercomputing Centre*
*Fabio Baruffa, PhD, Technical Consulting Engineer, Intel Corporation*

**Abstract**

As modern scientific simulations grow ever more in size and complexity, even their analysis and post-processing becomes increasingly demanding, calling for the use of HPC resources and methods. `yt` is a parallel, open source post-processing python package for numerical simulations in astrophysics, made popular by its cross-format compatibility, its active community of developers and its integration with several other professional python instruments. The **Intel® Distribution for Python** enhances `yt`'s performance and parallel scalability, through the optimization of lower-level libraries Numpy and Scipy, which make use of the optimized **Intel® Math Kernel Library** (**Intel-MKL**) and the **Intel MPI library** for distributed computing. The library package `yt` is used for several analysis tasks, including integration of derived quantities, volumetric rendering, 2D phase plots, cosmological halo analysis and production of synthetic X-ray observation. In this paper, we provide a brief tutorial for the installation of `yt` and the Intel distribution for python, and the execution of each analysis task.

Compared to the Anaconda python distribution, using the provided solution one can achieve net speedups up to **4.6x** on **Intel® Xeon Scalable processors** (codename Skylake).

## Installing `yt` on Intel and Anaconda python environment

We will need a python3 environment (we chose version 3.6) to create the conda package manager. For the installation of the Intel optimized software, one needs to add the Intel channel first (see [1]):

```
conda config --add channels intel
```

One can then create the environment, which we name `my_yt`, and install `yt` and all the packages we will need below, including the required dependencies. For Intel, the procedure is the following:

```
conda create -y -c intel -n my_yt python=3.6
source activate my_yt
conda install -y -c intel numpy scipy sympy mpi4py matplotlib
conda install -y -c conda-forge yt
conda install -y -c jzuhone -c astropy pyxsim
```

The repetition of the channel name ensures the correct origin of all dependency packages, while the `-y` option suppresses the confirmation prompt. Conda's installation help can be displayed at any time with `conda install -h`.

To use Anaconda python, which we need for the performance comparison, one just replaces `-c intel` by `-c anaconda` (and changes the environment name). Alternatively, one can use the all-in-one installation script from the `yt` webpage, providing the anaconda version within the miniconda environment.

## Tutorial: common post-processing tasks

We review a number of tasks which alone allow to analyze astrophysical simulations; these and more advanced options are available on the `yt` 3.5 tutorial [2]. We use a cosmological simulation including stars, dark matter and interstellar gas run with ENZO [3], but `yt` reads several formats, making it for instance quite a general-purpose volume-renderer.

To ensure proper `mpi4py` parallelization, the tasks must be scripted and python invoked in parallel, with *mpiexec* or equivalent (e.g. `mpiexec -np 8 python my_tasks.py`); timings for performance measurements can be added, e.g. with the python *time* package.

We begin by importing `yt`, enabling mpi4py, loading the chosen snapshot and performing a spherical selection `sp`, of radius 10 Megaparsec.

```python
import yt, mpi4py
yt.enable_parallelism()
ds = yt.load("RD0028/RedshiftOutput0028")  # Opening file header only
sp = ds.sphere("c", (10.,"Mpc"))           # Central 10 Megaparsec sphere
```

YT knows several **derived quantities** to compute from the snapshot datafields. We choose the total angular momentum per unit mass `j` of both gas and particles (dark matter, stars), interesting for halo shape studies, or matter accretion purposes (e.g. around Black Holes). This is calculated algebraically from 3D positions and velocities, and integrated over `sp` in just one line:

```python
j = sp.quantities.angular_momentum_vector(use_gas=True, use_particles=True)
```

and then `print(j)` prints the value of `j` in the correct unit system.

It is likewise easy to learn about the thermodynamic state of the gas through a **phase-plot**, binning the fields in 2D histograms. One can thus obtain pressure-volume diagrams for energy measures, or density-temperature diagrams indicative of the gas' *equation of state*.

```python
pp = yt.PhasePlot(sp, "density", "temperature", ["cell_mass"], weight_field=None)
pp.save()
```

The two, simply additive, tasks described above make use of `yt`'s efficient **Grid Parallelism**, i.e. a straightforward distribution among the active mpi4py processes of the basic elements (grids or particles) over which the fields are defined [4]. For more complex tasks such as **volume rendering** or **cosmological halo analysis** a proper **Spatial Decomposition** (more powerful but less efficient) of the domain is necessary instead. Volume renderings in `yt` are obtained via **ray-casting**, and in the most compact instance, an image `im` of the gas density is printed by:

```python
im, sc = yt.volume_render(ds, ('gas', 'density'), fname='volume.png')
```

although refined controls on the scene-object `sc` (transfer function, colormap, camera), and over the image itself (sigma-clipping, opacity) are available.

Cosmological halo analysis, i.e. compiling a catalogue of the properties of the formed structures, is a task which we also test, although a rather involved one (having to identify the halos in the first place, for which several methods exist in `yt`). Thus we point the interested reader to the `yt` documentation.

We finally showcase synthetic X-ray observations, via the `pyXSIM` package, using also `soxs` (dependency of `pyXSIM`) to simulate a real telescope (Chandra's ACIS-I)[5]. We begin by setting the parameters of telescope and observation, including a thermal X-ray emission model for the gas, telescope collecting area and the exposure time:

```
import soxs, pyxsim
soxs.soxs_cfg.set("soxs", "response_path", "./soxs_responses" )
redshift = 0.05
src_model = pyxsim.ThermalSourceModel("apec", 0.05, 11.0, 10000, Zmet=0.3)
exp_time = (100., "ks")
area     = (2000.0, "cm**2")
sp = ds.sphere("c", (50.,"Mpc")) # Enlarge to 50 Megaparsec
```

pyxsim will then compute individual photon packages via Montecarlo radiative transfer (**photons** task), and finally project them onto the detector (**events** task). The model includes hydrogen absorption.

```
# Computation 1/2: Montecarlo radiative transfer
photons = pyxsim.PhotonList.from_data_source(sp, redshift, area, exp_time, src_model)
# Computation 2/2: Photons produce events into simulated detector
events_z = photons.project_photons("z", (45.,30.), absorb_model="tbabs", nH=0.04)
```

The data can then be printed out, and the simulated instrument applied (optional); exposure, celestial coordinates  (here 45 and 30 degrees) and photon energy extrema (here set to 0.5 and 11 electronVolt) can be adjusted. This completes all scripted tasks.

```
events_z.write_simput_file("RD0028", overwrite=True) # Warning: very large fits file!
soxs.instrument_simulator("RD0028_simput.fits", "evt.fits", (100.0, "ks"), "acisi_cy0",
[45., 30.], overwrite=True)
soxs.write_image("evt.fits", "img.fits", emin=0.5, emax=11.0, overwrite=True)
exit()
```

**Speedup achieved by Intel Pyhton**

We run a scaling test of all the tasks for both Anaconda and Intel python over single **SKX Platinum 8174** nodes, and plot the shortest execution time (median of 20 measurements) in Figure 1 below.

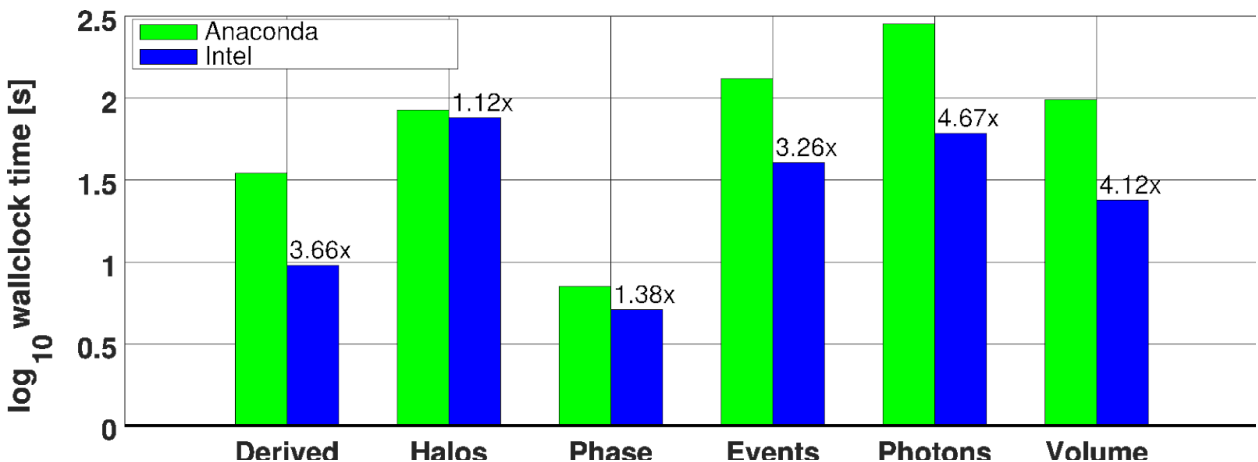

**Fig.1 Performance comparison of Anaconda and Intel python on all tasks. We plot the shortest execution time (log scale) on one SKX node. Intel python improves the performance up to 4.6x.**

Intel python improves the **performance** up to a factor 4.6x. Longer tasks tend to show larger improvements; the **halos** taks is an exception, as instead of grid or spatial parallelism yt distributes individual halos (once found) among the processes:  the full datafields are still accessed for computation, so the task is not embarrassingly parallel, but the work-sharing is easier. The observed speedup is nonetheless due to a better scaling; the value may increase with the number of halos.

Concerning **code scalability**, Anaconda performed always better in serial than in parallel, except for the **events** task, scaling up to two `mpi4py` processes. So the only convenient parallelization on Anaconda is unfortunately `yt`'s embarassingly parallel scheme over time series or different objects [4]. Intel python instead scales easily up to 8 or 16 cores, allowing a much better usage of the shared resources, highly valuable in the case of larger tasks or simulations.